\documentclass{PoS}
\PoS{PoS(LAT2005)175}
\usepackage{epsfig}
\usepackage{graphicx}
\usepackage{bm}

\def\Eq#1{Eq.~(\ref{#1})}
\def\eff{\rm eff}

\def\<{\langle}
\def\>{\rangle}

\title{The pressure and a possible hidden Hagedorn transition at large-N}

\ShortTitle{The pressure and a possible hidden Hagedorn transition at large-N}

\author{\speaker{B.~Bringoltz} and M.~Teper\\
        Rudolf Peierls Centre for Theoretical Physics,\\
        University  of  Oxford,\\ 
        1 Keble Road, Oxford, OX1 3NP, UK\\
        E-mail: \email{barak@thphys.ox.ac.uk},
                \email{m.teper1@physics.oxford.ac.uk}}

\abstract{In the first part of this contribution we present a numerical study motivated by recent attempts to understand the nonperturbative aspects of QCD at temperatures $T\sim$ a few times the deconfinement temperature $T_c$. We focus on the pure gauge theory, and ask whether the deficit in pressure and entropy, with respect to their free-gas values, is particular to $SU(3)$. We find that the deficit in $SU(4),SU(8)$ for $T\le 2T_c,1.6 T_c$, respectively, is remarkably close to that of $SU(3)$. This suggests a similar deficit for $SU(\infty)$, which is fortunate since this theory is simpler, and can serve to constrain the possible dynamics underlying the deficits. 

In the second part we seek for signs of a Hagedorn temperature $T_H$ in pure lattice $SU(N)$ gauge theories with $N=8,10,12$. Since one expects $T_H>T_c$, we measure masses of strings in the metastable confined phase above $T_c$, and extrapolate them to zero to estimate $T_H$. For $SU(12)$ we find that $T_H/T_c=1.116(9)$, when we extrapolate with a critical exponent of the three dimensional XY model, which seems to be preferred over a mean field exponent by our data.}

\FullConference{XXIIIrd International Symposium on Lattice Field Theory\\
                 25-30 July 2005\\
                 Trinity College, Dublin, Ireland}

\begin{document}

\section{Bulk thermodynamics of lattice $SU(N)$ gauge theories at large-$N$}
\label{sec:pressure}

In this section we present a study of bulk thermodynamical properties of $SU(4)$, and $SU(8)$ pure lattice gauge theories and compare to a study by Boyd et al. of $SU(3)$ \cite{Boyd_et_al}. Recent calculations 
of various  properties of $SU(N)$ gauge theories 
\cite{Lucini:2002ku}--\cite{Lucini:2004my}
have demonstrated that $N=8$ is very close to $N=\infty$ 
for most purposes. These also provide information on the critical coupling for various 
lattice sizes and $N$.
Thus our calculations should provide us with an accurate picture 
of what happens to bulk thermodynamics at $N=\infty$. For a more detailed version of this section we refer to \cite{pressure_paper}.

\subsection{Lattice setup and methodology}
\label{sec:lattice}

We define the gauge theory on a discretised periodic Euclidean four 
dimensional space-time with $L^3_s\times L_t$ sites, and perform Monte-Carlo simulations of a simple Wilson action.
We use the Kennedy-Pendelton heat bath algorithm for the link 
updates, followed by five over-relaxations of all the $SU(2)$ 
subgroups of $SU(N)$. To evaluate bulk thermodynamics we choose to use the ``integral method'' (see for example \cite{Boyd_et_al})\footnote{Our lattices are too coarse for the ``differential'' method, and we found that the Wang-Landau algorithm for the evaluation of the density of states, did not converge easily for $SU(8)$.} and so the pressure $p$ and interaction measure $\Delta$ are
\begin{equation}
p/T^4
=
6 L^4_t \int^\beta_{\beta_0} d\beta^\prime 
(\langle u_p \rangle_T - \langle u_p \rangle_0), \qquad
\label{eq:pint2} 
\Delta/T^4
= 6 L_t^4 
(\langle u_p(\beta) \rangle_0 - \langle u_p(\beta) \rangle_T)
\times
\frac{\partial \beta}{\partial \log (a(\beta))}. 
\label{eq:final_delta}
\end{equation}
Here $\langle u_p \rangle_{T}$ is the plaquette average on a $T> 0$ lattice with $L_t<L_s$, while $\langle u_p \rangle_{0}$ is measured on a lattice with relatively large $L_t=L_s$. To evaluate the scaling of the temperature $T$ with the coupling $\beta=2N/g^2$, and the derivative in \Eq{eq:final_delta}, we use 
calculations of the string tension, $\sigma$,
in lattice units (e.g. \cite{Lucini:2005vg}). 
Other physical choices to fix the scale differ in modest $O(a^2)$ differences. Finally the value of $T_c/\surd\sigma$ for the different gauge groups and $L_t=5$ in taken from \cite{Lucini:2003zr,Lucini:2005vg}.

We performed calculations of $\langle u_p \rangle_T$ in $SU(4)$ on $16^3 5$ lattices and 
in $SU(8)$ on $8^3 5$ lattices for a range of 
$\beta$ values corresponding to $T/T_c \in [0.89,1.98]$ for 
$SU(4)$, and to $T/T_c\in [0.97,1.57]$ for $SU(8)$.
Since we use 
$L_t=5$, while the data for $SU(3)$ in \cite{Boyd_et_al} is 
for $L_t=4,6,8$, we also performed simulations for $SU(3)$ 
on $20^3 5$ lattices with $T/T_c\in [1,2]$.
The measured plaquette averages are presented in \cite{pressure_paper}.

We performed the `$T=0$' calculations of $\langle u_p\rangle_0$
on $20^4$ lattices for SU(3), and on
$16^4$ lattices for SU(4), which are known to be effectively at $T=0$ 
\cite{Lucini:2001ej,Lucini:2004my} for the couplings involved.
For SU(8) 
however, using $8^4$ lattices would not be adequate for the
largest $\beta$-values, and we 
take instead the SU(8) calculations on larger 
lattices in
\cite{Lucini:2004my},
and interpolate between the values of $\beta$ used there with the 
ansatz
$\langle u_p \rangle_0(\beta)
=\langle u_p \rangle^{P.T.}_0(\beta)
+\frac{\pi^2}{12}\frac{G_2}{N\sigma^2}(a\sqrt{\sigma})^4
+c_4g^8+c_5g^{10}, $
where $\langle u_p \rangle^{P.T.}_0(\beta)$ is the lattice 
perturbative result to ${\cal O}(g^6)$ from 
\cite{Alles:1998is}
and $N=8$.
Our best fit has $\chi^2/{\rm dof}=0.93$ with ${\rm dof}=2$,
and the best fit parameters are $c_4=-6.92$, $c_5=26.15$, and a 
gluon condensate of $\frac{G_2}{N\sigma^2}= 0.72$. 

\subsection{Finite volume effects}
For $N=4,8$, one is able to use lattice volumes much
smaller than what one needs for $SU(3)$ (like those in \cite{Boyd_et_al}) as the longest correlation length decreases rapidly with $N$
\cite{Lucini:2003zr,Lucini:2005vg}. This is also theoretically expected, much more generally, as $N\to\infty$. 
The main remaining concern has to do with tunneling configurations, which occur 
only at $T_c$ when $V\to\infty$.
On our finite volumes, this is no longer true, and we minimise finite-$V$ corrections
by calculating the average plaquettes only in field configurations
that are confining, for $T<T_c$, or deconfining, for $T>T_c$.
For SU(3), where the phase transition 
is only weakly first order, it is not
practical to attempt to separate phases. This will smear the apparent variation of the pressure across $T_c$ in the
case of SU(3).

To confirm that our finite volume effects are under control we have compared the SU(8) value of $\langle u_p(\beta) \rangle$ as
measured in the deconfined phase of the our $8^3\times 5$ lattice 
with other $L_s^3\times 5$ results from other studies 
\cite{Hagedorn_paper}, and find that the results are consistent
at the $2$ sigma level.
We perform a similar check for the confined phase on the same lattices, and again find that finite volume effects are small, mostly on a one sigma level. The data supporting these checks, together with the checks related to the next paragraph, is presented in \cite{pressure_paper}.

A similar check for the confined phase on $L^4$ lattices leads to conclude that 
a size $L=8$ in SU(8) is not 
large enough, as we find that the plaquette average has a significant change (on a $16$ sigma level) from $L=8$ to $L=16$ for our largest value of $\beta$. 
By contrast, for $SU(4)$ the finite volume effects seem not to be
large on the $16^4$ lattice as we checked for our largest value of
$\beta=11.30$. There the value of the plaquette on a $20^4$ lattice is
consistent within $\sim 2.3$ sigma with the value on a $16^4$ lattice.

\subsection{Results}
\label{results}

In presenting our results for the pressure, we shall 
normalize to the lattice Stephan-Boltzmann result given by
$\left( p/T^4
\right)_{\rm{free-gas}}=(N^2-1)\frac{\pi^2}{45}\times R_I(L_t). $
Here $R_I$ includes the effects of discretization errors in the integral method \cite{Engels:1999tk,pressure_paper}. The same normalisation is applied for the interaction measure $\Delta$, and for the energy density $\epsilon=\Delta+3p$, and entropy density $s=(\Delta+4p)/T$.

\begin{figure}[htb]
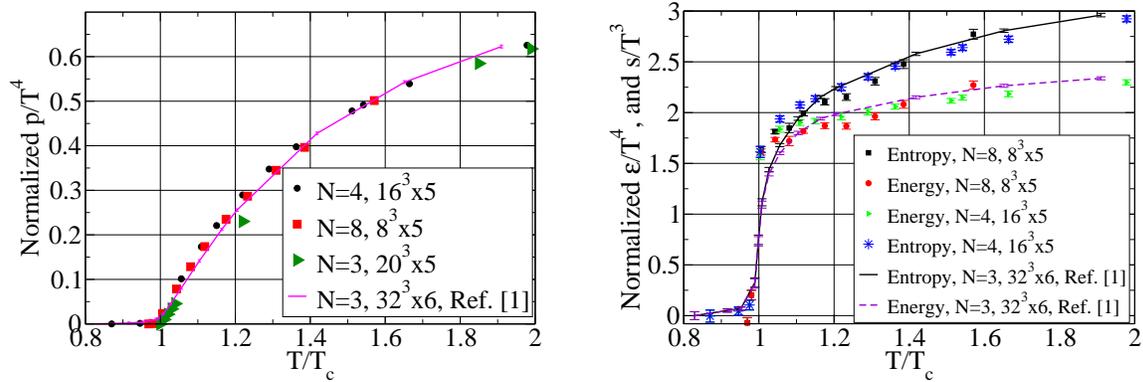

\begin{center}
\includegraphics[width=7.1cm,clip]{P348_lat05.eps} \qquad \includegraphics[width=7.1cm,clip]{EandS348_lat05.eps}\caption{The pressure (on the left), and energy and entropy densities (on the right) normalized to the lattice Stephan-Boltzmann 
pressure, including the full discretization errors given by $R_I$. In the pressure plot, the symbols' vertical sizes are representing the largest error bars (which are received for the highest temperature). The 
lines are the $SU(3)$ results and $L_t=6$ from 
\cite{Boyd_et_al}.} \label{fig1}
\vskip -0.5cm
\end{center}
\end{figure}

We present our $N=4$ and  $N=8$ results for $p/T^4$ in the left plot of 
Fig.~\ref{fig1}. We also show our calculations of the $SU(3)$
pressure for $L_t=5$, as well as the $L_t=6$ calculations from 
\cite{Boyd_et_al}. In the right plot of Fig.~\ref{fig1} we present results for the normalized 
energy density $\epsilon$, and normalized entropy density $s$. The lines are the $SU(3)$ result of 
\cite{Boyd_et_al} 
with $L_t=6$. Again we see very little dependence on the gauge group,
implying very similar curves for $N=\infty$.
One can clearly infer that the pressure 
in the $SU(4)$ and $SU(8)$ cases is remarkably close to that
in $SU(3)$ and hence that the well-known pressure deficit 
observed in SU(3) is in fact a property of the large-$N$ 
planar theory. This implies that the dynamics that drives the deconfined 
system far from its noninteracting gluon gas limit, must remain
equally important in the $N=\infty$ planar theory. This is encouraging 
since that limit is simpler to approach analytically, for example using gravity duals, and also because it can serve to constraint and point to important ingredients of analytical approaches (see the discussion in \cite{pressure_paper}).

\section{In search of a Hagedorn transition in $SU(N)$ lattice gauge theories at large-$N$}
\label{sec:Hagedorn}

In this section we present a calculation of the change in the mass $m(T)$ of Polyakov loops in pure lattice $SU(N)$ gauge theories with $N=8,10,12$, as the temperature $T$ crosses $T_c$. At these large values of $N$, interactions between singlet objects in the confined phase are suppressed, and a Hagedorn picture of string proliferation is most attractive. Because the strings proliferate with an effective zero mass,  one expect $m(T)$ to vanish at the Hagedorn temperature $T_H>T_c$, which we aim to estimate.

\subsection{Methodology and lattice calculation}

Although for $N\ge 3$ the first order deconfining transition 
occurs when $m(T_c)>0$ (and therefore $T_c < T_H$), the Hagedorn
transition may still be accessible. This is the case since the deconfining transition
is strongly first order at larger $N$, and one can try to use its metastability to
carry out calculations in the confining phase for $T> T_c$. 
The way $m(T)$ drops with $T$ tells us how does the curvature of the confining minimum in the effective loop potential vanishes. The simplest possibility is that the point where it vanishes is the spinodal point of the potential, where the confined vacuum becomes unstable (and where the barrier between the confining and deconfining vacua disappears). It is, however, quite possible that $T_H$
does not coincide with the spinodal point, and if the latter occurs below $T_H$, then it may (but need not) interfere with our 
determination of $T_H$. This is a significant ambiguity that we 
cannot resolve in the present calculations but the reader should
be aware of its existence. 

We therefore begin deep in the confined phase 
and increase $T$ to $T>T_c$,
calculating the decrease in $m(T)$, and extrapolating 
to $m=0$. We interpret the result of the extrapolation 
as the Hagedorn temperature, $T_H$. Nevertheless, since we work with 
finite values of $N$ and volume $V$, tunneling probably occurs 
somewhere below $T_H$. These tunneling effects and the fact that 
as $m(T)$ decreases, finite volume effects become 
important, can make an apriori fit for the critical exponents 
unreliable. As a result we first perform fits where we fix the 
functional behaviour of the temporal loop mass to be 
$m(T)=A\cdot \left( T_H/T_c-T/T_c \right)^\nu$, 
where $\nu=0.6715(3)$ corresponding to three dimensional XY model (3DXY) or $\nu=0.5$ corresponding to mean field. In addition we also perform fits 
where the exponent $\nu$ is a free parameter, constrained to be positive and smaller than one.

Our lattice calculations practicalities are the same as those in Section~\ref{sec:pressure}. We work with $L_t=5$, and $L_s=12$. In addition, every five sweeps, we measure the correlations functions of improved operators for Polyakov lines. The masses are extracted with a variational technique, and the results are given in \cite{Hagedorn_paper}, where the full detailed description of our lattice study (including statistics, thermalization details, choices of initial configuration, monte-carlo results, etc.) is given as well.

The physical scale was fixed 
using the interpolation for the string
tension given in \cite{Lucini:2005vg} in the case of $SU(8)$. For
$N=10,12$ we extrapolate the parameters of the scaling function of
\cite{Lucini:2005vg} in $1/N^2$ from their values at $N=6,8$. In addition we
measured the string tension for $N=10$ at $\beta=68.80$
on an $8^4$ lattice, and for $N=12$ at $\beta=99.2,100.0$, on an $8^4$, and a $10^4$ lattice respectively. We find that the measured values deviate at most by $1.6$ sigma from the calculated ones.

Finally, to check the effect of finite volume corrections, we perform several
additional calculations of the correlation lengths for $SU(10)$ with $L_s=14$, and for $SU(12)$ with $L_s=16$. We find that the finite volume effects of the extracted masses are quite small, and are at most on the level of $1.9$ sigma (and mostly lower than that, see \cite{Hagedorn_paper}). Comparing
with the situation close to the second order phase transition of the
$SU(2)$ group \cite{Lucini:2005vg}, we find that for $N=2$, finite
volume effects are much more important than for $N=12$, which is consistent with standard theoretical arguments that predict smaller volume corrections for gauge theories with larger values of $N$.

We estimated $T/T_c$ by $\frac{(a\sqrt{\sigma})_c}{a\sqrt{\sigma}}$, where $(a\sqrt{\sigma})_c$ is the 
value of $a\sqrt{\sigma}$ at $\beta=\beta_c$ on an $L_t=5$ lattice\footnote{This assumes that  $T_c/\sqrt{\sigma}$ varies at most
very weakly with $a(\beta)$, which is in fact what one observes \cite{Lucini:2005vg}.}. The latter we extract from 
$T_c/\sqrt{\sigma}$, which is 0.5819(41) for
$N=8$.
The corresponding value for the
$N=10,12$ is found by extrapolating in $1/N^2$ according to measured
values of $T_c/\sqrt{\sigma}(L_t=5)$ for $N=4,6,8$. This gives 0.5758 for $SU(10)$ and 0.5735 for
$SU(12)$, with an error of about $1\%$.

\subsection{Results}

\begin{figure}[htb]
\begin{center}
\includegraphics[width=7cm,clip]{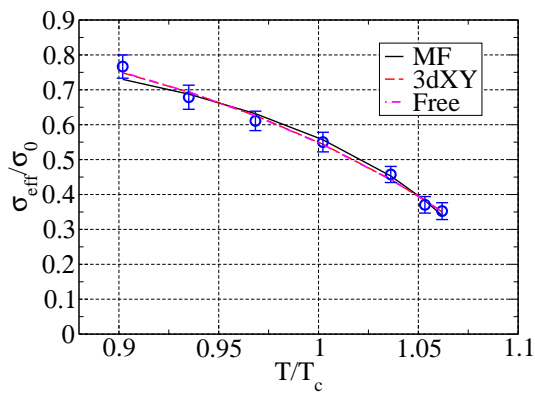}
\caption{Effective string tension for $SU(12)$ obtained in units of the zero temperature string tension.} \label{fig:N12}
\end{center}
\vskip -0.8cm
\end{figure}

Here we focus only on the results of the $SU(12)$ study. We choose to plot the effective string tension $\sigma_{\eff}(T)\equiv m(T)\, T$, in units of the zero temperature string tension $\sigma_0$, in Fig.~\ref{fig:N12}.

We find that the fits have a rather good $\chi^2/{\rm dof}=0.30,0.50$ when fitting with $\nu=0.6715,0.5$, and give $T_H/T_c=1.116(9)-1.092(6)$, respectively. The result of the fit with $\nu$ as a free parameter is $\nu_{\rm free}=0.69$, and $T_H/T_c=1.119$, with $\chi^2/{\rm dof}=0.37$ (the error here is correlated with the error of $\nu$ and $A$. (For the iso-surfaces that correspond to confidence levels see \cite{Hagedorn_paper}). With this weak preference of $\nu=0.6715$ we cannot unambiguously determine the exponent $\nu$ (the error on $\nu_{\rm free}$ is quite large, which can be seen from the goodness of all these three fits). Nonetheless, in the case of $SU(10)$ we find a stronger preference for $\nu=0.6715$, see \cite{Hagedorn_paper}.

The limited statistics prevent us from making statements about the function $T_H(N)$. This is unfortunate, since it
is of interest to know how far is $T_H/T_c$ from $1$ at
$N=\infty$. Nevertheless for $N=8,10,12$ we obtain fitted values of
$T_H/\sqrt{\sigma}\simeq 0.62-0.68$, which is lower than
$T_c/\sqrt{\sigma}\simeq 0.7$ of $SU(2)$ where
the phase transition is second order, and therefore may be Hagedorn,
$T_c=T_H$, or provides a lower bound on $T_H$. Nonetheless, when we plot the effective string tension for $N=10,12$ as a function of $T/\surd \sigma$ we find that they are quite similar, but differ from the plot of $\sigma_{\eff}(T)$ of $N=2$, which may question the identification of deconfinement at $N=2$ with a Hagedorn transition. Finally, an additional outcome of this work 
is to confirm that  at $T=T_c$ the mass of the timelike flux loop 
that couples to Polyakov loops is far 
from zero at large-$N$ ($\sigma_{\eff}/\sigma_0\simeq 0.5$ for $N=8,10,12$, as seen in Fig.~\ref{fig:N12} here and in \cite{Hagedorn_paper}), which confirms that the transition is indeed strongly first-order.

\begin{acknowledgments}

We are thankful to J.~Engels for discussions on the discretisation errors of the free lattice gas pressure, and for giving us the numerical routines to calculate them. We also thank the conference organizers for the opportunity to present these works.

\end{acknowledgments}


\begin{thebibliography}{99}

\bibitem{Boyd_et_al}
G.~Boyd, J.~Engels, F.~Karsch, E.~Laermann, C.~Legeland, M.~Lutgemeier and B.~Petersson, \emph{Thermodynamics of SU(3) Lattice Gauge Theory}, Nucl.\ Phys.\ B {\bf 469}, 419 (1996) [arXiv:hep-lat/9602007].

\bibitem{Lucini:2002ku}
B.~Lucini, M.~Teper and U.~Wenger, \emph{The deconfinement transition in SU(N) gauge theories}, Phys.\ Lett.\ B {\bf 545}, 197 (2002) [arXiv:hep-lat/0206029].

\bibitem{Lucini:2003zr}
B.~Lucini, M.~Teper and U.~Wenger, \emph{The high temperature phase transition in SU(N) gauge theories}, JHEP {\bf 0401}, 061 (2004) [arXiv:hep-lat/0307017].

\bibitem{Lucini:2005vg}
B.~Lucini, M.~Teper and U.~Wenger, \emph{Properties of the deconfining phase transition in SU(N) gauge theories}, JHEP {\bf 0502}, 033 (2005) [arXiv:hep-lat/0502003].

\bibitem{Lucini:2001ej}
B.~Lucini and M.~Teper, \emph{SU(N) gauge theories in four dimensions: Exploring the approach to N = infinity}, JHEP {\bf 0106}, 050 (2001) [arXiv:hep-lat/0103027].

\bibitem{Lucini:2004my}
B.~Lucini, M.~Teper and U.~Wenger, \emph{Glueballs and k-strings in SU(N) gauge theories: Calculations with  improved operators}, JHEP {\bf 0406}, 012 (2004) [arXiv:hep-lat/0404008].

\bibitem{pressure_paper}
B.~Bringoltz and M.~Teper, \emph{The pressure of the SU(N) lattice gauge theory at large-N}, arXiv:hep-lat/0506034. 

\bibitem{Alles:1998is}
B.~Alles, A.~Feo and H.~Panagopoulos, \emph{Asymptotic scaling corrections in QCD with Wilson fermions from the  3-loop average plaquette}, Phys.\ Lett.\ B {\bf 426}, 361 (1998) [Erratum-ibid.\ B {\bf 553}, 337 (2003)] [arXiv:hep-lat/9801003].

\bibitem{Engels:1999tk}
J.~Engels, F.~Karsch and T.~Scheideler, \emph{Determination of anisotropy coefficients for SU(3) gauge actions from  the integral and matching methods}, Nucl.\ Phys.\ B {\bf 564}, 303 (2000) [arXiv:hep-lat/9905002].
J.~Engels,
private Communications (2005)

\bibitem{Hagedorn:1965st}
R.~Hagedorn, \emph{Statistical Thermodynamics Of Strong Interactions At High-Energies}, Nuovo Cim.\ Suppl.\  {\bf 3}, 147 (1965). 

\bibitem{Hagedorn_paper}
B.~Bringoltz and M.~Teper, \emph{In search of a Hagedorn transition in SU(N) lattice gauge theories at large-N}, arXiv:hep-lat/0508021.

\end{thebibliography}
\end{document}